\documentstyle[preprint,aps,epsfig]{revtex}

\begin{document}

\title{Statistical Exploration of Fragmentation Phase Space and \\
Source Sizes in Nuclear Multifragmentation}

\author{L.~G.~Moretto, L.~Beaulieu$^*$, L.~Phair, and G.~J.~ Wozniak}

\address{Nuclear Science Division, Lawrence Berkeley National
Laboratory, Berkeley, California 94720}

\date{\today}

\maketitle

\begin{abstract}
The multiplicity distributions for individual fragment $Z$ values in
nuclear multifragmentation
are binomial. The extracted maximum value of the multiplicity, $m_Z$,
is found to depend on $Z$ according to $m_Z=Z_0/Z$, where $Z_0$ is the
source size. This is shown to be a strong indication of statistical
coverage of fragmentation phase space. The inferred source sizes
coincide with those extracted from the analysis of fixed multiplicity
charge distributions.
\end{abstract}

\pacs{25.70.Pq}

\narrowtext

Is nuclear multifragmentation \cite{Moretto93,Bon95} a statistical or
a dynamical process?  Let us consider a fragmentation space defining
the number and masses/charges of the produced fragments. The thermal
population of each cell can be expressed in terms of suitable
Boltzmann factors. The thermal nature of this population can be tested
in a variety of ways. A very visual way is the Arrhenius plot where
the log of the population is plotted vs.~the reciprocal temperature
$1/T$.

The question however, remains whether, apart from the Boltzmann
factor, all the cells of the fragmentation space are uniformly
explored. A test of such a uniform filling implies the knowledge of
the size (mass/charge) of the source.
A new empirical feature observed in many reactions has led us to
a way to verify the uniform filling of fragmentation space and to
determine simultaneously the source size.

It has been shown \cite{Moretto97} that the intermediate mass fragment
(IMF) multiplicity distribution $P_n$ at any given transverse energy
$E_t$ is {\it empirically} given by a binomial distribution
\begin {equation}
P_n = \frac{m!}{n!(m-n)!} p^n (1-p)^{m-n}.
\label{eq:binomial}
\end {equation}
This
implies that fragments are emitted nearly independently of each other,
so that the probability $P_n$ of observing $n$ fragments can be
written by combining a single one-fragment emission probability $p$
according to Eq.~(\ref{eq:binomial}).  The parameter $m$ (the total
number of throws) represents the maximum possible number of fragments,
which is immediately related to the source size.

    The simplest statistical equilibrium model of multifragmentation
has exactly the structure of Eq.~(\ref{eq:binomial}).  Let us assume
that the source is made up of $m$ fragments. The ``outside'' fragments
have energy $\epsilon_2$, and those ``inside'' have energy
$\epsilon_1$.  A generic partition of $n$ fragments outside and $m -
n$ fragments inside has the probability:
\begin {equation}
P_n = \frac{m!}{n!(m-n)!} \frac {e^{-(n\epsilon_2 +
(m-n)\epsilon_1)/T}} {(e^{-\epsilon_1/T} + e^{-\epsilon_2/T})^m}
\label{eq:two_level}
\end {equation}
which leads to Eq.~(\ref{eq:binomial}) when
\begin {equation}
p = \frac{e^{-\epsilon_2/T}}{e^{-\epsilon_1/T} + e^{-\epsilon_2/T}}.
\label{eq:number_4}
\end {equation}
Thus, a simple way to obtain the size of the source is to multiply $m$
by the fragment size.

When the definition of IMF covers a range of atomic numbers
(IMF: $Z_{th}\le Z\le 20$, with $Z_{th}$
equal to 3), one should multiply $m$ by a ``suitably'' averaged $Z$.
In fact, a dependence of $m$ on the lower threshold $Z_{th}$ has been
found such that $m(Z_{th})\times Z_{th} \approx constant$
\cite{Moretto97}.  The natural next step is to restrict the fragment
definition to a single atomic number $Z$.

A straightforward generalization of Eq.~(\ref{eq:binomial}) to the
production of fragments with charges 1, 2, ...$Z_0$ is given by the
multinomial distribution
\begin{equation}
P=\frac{Z_0!}{n_1!n_2!...n_{Z_0}!}
p_1^{n_1}p_2^{n_2}...p_{Z_0}^{n_{Z_0}}
\label{eq:multinomial}
\end{equation}
with
\begin{equation}
Z_0=\sum _ZZn_Z.
\label{eq:sum}
\end{equation}
Here there is no single quantity $m$ as in Eq.~(\ref{eq:binomial}),
since the constraint is now on the total charge rather than on the
total number of fragments.
However, this leads immediately to a simple scaling that must be
obeyed if the fragmentation phase space is to be completely explored.

Let us first consider, as a familiar example, the fragmentation space
spanned by the Euler number partitions. This case is particularly
simple because all partitions are unbiased, or equally probable. This
would correspond, for the different model defined in
Eqs.~(\ref{eq:multinomial}) and (\ref{eq:sum}), to setting
$p_i$=constant.

Let us indicate with $W(Z_0)$ the statistical weight (number of
partitions) associated with the integer or ``charge'' $Z_0$. If we now
select the partitions containing at least $n_Z$ integers of size $Z$,
their number
is given by the number of partitions of the residue
\begin{equation}
P_{n_Z}= W(Z_0-n_ZZ)
\end{equation}
by definition.  Since the ``charge'' constraint is applied
``minimally,'' what counts is only the product $Zn_Z$, rather than the
individual $Z$ values.  Consequently, the weight of the residue does
not change if we substitute $n_ZZ$ with $n_{Z'}Z'$ provided that
$n_ZZ=n_{Z'}Z'$.
In other words,
$P_{n_Z,Z}= P_{n_{Z'},Z'}$
if $n_ZZ=n_{Z'}Z'$.


This gives immediately the scaling laws
\begin{equation}
n_Z/n_{Z'}=Z'/Z
\end{equation}
or for the extreme value of $n_Z=m_Z$,
\begin{equation}
m_Z=Z_0/Z.
\label{eq:m_Zscaling}
\end{equation} 
This
result follows directly from the uniform exploration of the
fragmentation phase space.
It amounts to a {\em complete decoupling} of all the fragmentation
paths. Once the system attains the charge loss of $q=nZ$, it is
indifferent to how this was attained. {\em Any} path $q=\sum n_iZ_i$
is {\em equivalent} and {\em substitutable}. Similarly, for the
remaining charge $Z_0-q$.

This argument works when all
partitions are unbiased, as for instance in the Euler number partition
mentioned above. However, in the binomial/multinomial distributions of
Eqs.~(\ref{eq:binomial}) and (\ref{eq:multinomial}), each partition is
weighted by the probabilities $p_i$ (reflecting $Q$-value effects)
which may or may not be Boltzmann factors.

Yet, the binomial/multinomial analysis elegantly and automatically
separates out the $Q$-value dependent probability $p$ from the
parameter $m_Z$ which now contains direct information on the
accessible fragmentation space and must follow the scaling given by
Eq.~(\ref{eq:m_Zscaling}). Thus the $1/Z$ scaling is the counterpart
of the Arrhenius plot which verifies the thermal (Canonical)
population of each fragmentation configuration.

When the restriction to {\em individual} $Z$ values is made
experimentally \cite{Beaulieu98}, the multiplicity distributions are
found to be nearly Poissonian, namely $mp << m$. This introduces
interesting simplifications in the analysis and interpretation of the
data, but at the cost of a loss of scale. In the Poisson limit the
average multiplicity $\langle n \rangle = mp$ is the only accessible
parameter, and the decomposition into $m$ and $p$ becomes impossible.
The recovery of scale for an individual $Z$ is highly desirable in
view of the possibility that the number of throws $m_Z$ (for a single
species $Z$) might obey the simple scaling of
Eq.~(\ref{eq:m_Zscaling}).

    Thus, we have attempted binomial fits of the multiplicity
distributions for {\it individual} $Z$ values in an effort to extract
$m_Z$.  Fortunately, a number of reactions ($^{36}$Ar+$^{197}$Au at 35
to 110 $A$MeV \cite{Des91}, $^{129}$Xe+$^{27}$Al,$^{51}$V,$^{\rm
nat}$Cu, $^{89}$Y, at 50 $A$MeV \cite{Bow92}, and
$^{129}$Xe+$^{197}$Au at 50 to 60 $A$MeV \cite{Tso95}) have been
studied with good $Z$ resolution and high statistics.
We first consider the asymmetric, intermediate-energy reactions in
reverse kinematics exemplified by $^{129}$Xe+$^{\rm nat}$Cu
at 50 $A$MeV, for which we can expect a single dominant fragment
source.

\begin{figure}
\centerline{\psfig{file=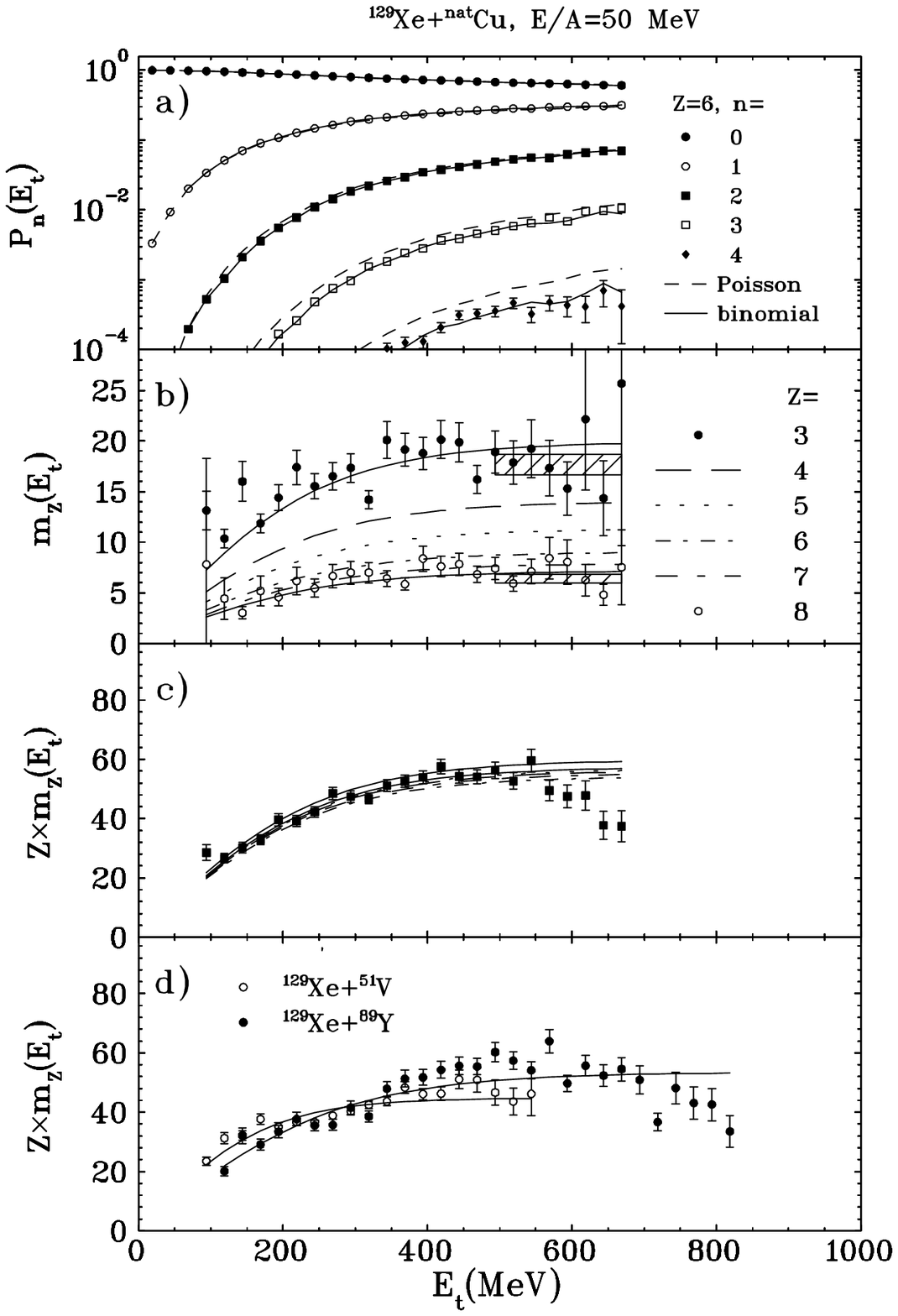,height=12.5cm,angle=0}}
\vspace{0.2cm}
\caption{Panel a): the $n$-fold probability distribution (symbols) and
Poisson (dashed) and binomial (solid) fits are plotted as a function
of transverse energy for carbon fragments emitted from the reaction
$^{129}$Xe+$^{\rm nat}$Cu at 50 $A$MeV.  Panel b): the extracted
binomial parameter $m_Z$ (the number of ``throws'') is plotted as a
function of transverse energy for lithium (solid circles) and oxygen
(open circles) emission. The solid lines are hyperbolic tangent fits
to the indicated data. The other
lines are fits to data not shown. The two hatched regions represent
weighted averages of the top 5\% most central collisions (based on the
integrated $E_t$ spectrum) of the $m_Z$ values for lithium and
oxygen. Panel c): The fits from panel b) are scaled by the atomic
number $Z$ of the emitted particle.  The square symbols represent an
``average'' source size calculated with
Eq.~(\protect\ref{eq:average_size}).  Panel d): The symbols represent
an ``average'' source size calculated with
Eq.~(\protect\ref{eq:average_size}) for $^{129}$Xe+$^{51}$V and
$^{129}$Xe+$^{89}$Y. The lines are hyperbolic tangent fits to the data.  }
\label{fig:Pn}
\end{figure}

Examples of both binomial and Poisson fits to the carbon yield from
this reaction are shown in panel a) of Fig.~\ref{fig:Pn}. An
improvement of the fit by using the binomial expression is observed
for large fold numbers. A similar improvement is observed for each $Z$
in all reactions listed in this letter.

The $E_t$ dependence of the parameters $m_Z$ from the binomial fits to
the multiplicity distributions associated with each fragment atomic
number leads to several observations.  For each $Z$ value, $m_Z$
increases to a near constant value with increasing $E_t$. We
approximate this behavior with
the form
$m_Z=m_Z^0\tanh fE_t$.
The parameter $m_Z^0$ represents the saturation value of $m_Z$ for
large $E_t$ and $f$ controls the rise of $m_Z$ with increasing $E_t$.
The solid lines in panel b) of Fig.~\ref{fig:Pn} are the empirical
fits to $m_Z$ values extracted for lithium and oxygen emission from
the reaction $^{129}$Xe+$^{\rm nat}$Cu at 50 $A$MeV. The other
discontinuous lines are fits to data not shown ($Z$=4-7).

\begin{figure}
\centerline{\psfig{file=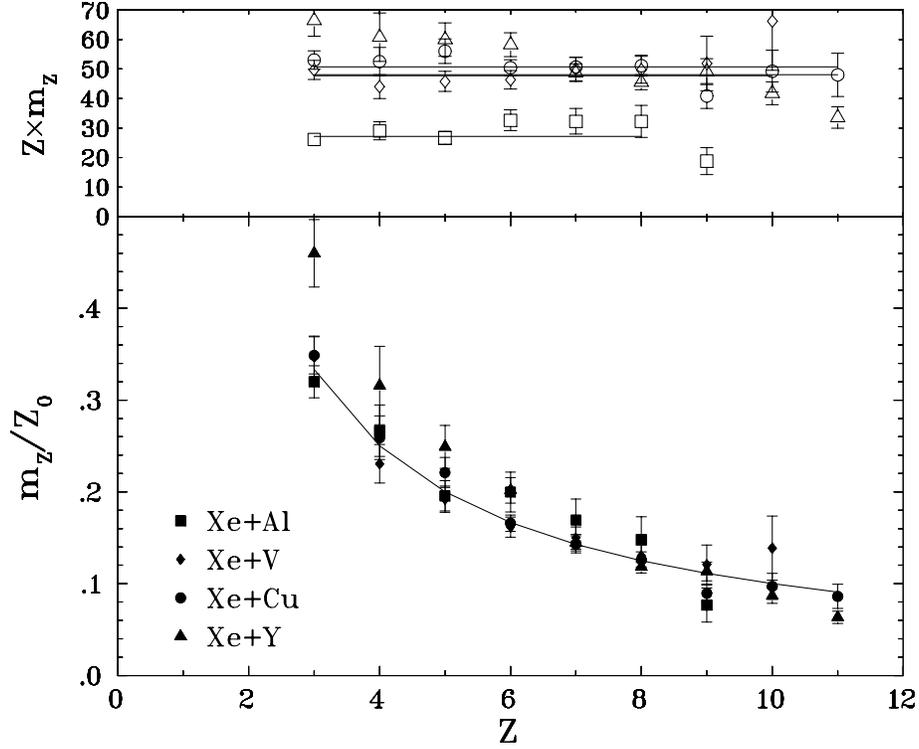,height=12.0cm,angle=90}}
\vspace{0.2cm}
\caption{Top panel: The $m_Z$ values scaled by $Z$ as a function of
$Z$ for the indicated reactions (open symbols) at 50 $A$MeV extracted
for the 5\% most central collisions. Bottom panel: The values of $m_Z$
divided by the value of the extracted source charge $Z_0$ for the
indicated reactions (solid symbols). The solid line is $1/Z$.  }
\label{fig:mZ0_xecu}
\end{figure}

Furthermore, at all $E_t$ values there is an overall decrease of $m_Z$
with increasing fragment $Z$ value in agreement with the expected
scaling $Zm_Z = Z_0$.  This remarkable dependence is exemplified in
panel c) of Fig.~\ref{fig:Pn} and in Fig.~\ref{fig:mZ0_xecu}.  By
applying the expected scaling ($Zm_Z$), all of the fits to the
$^{129}$Xe+$^{\rm nat}$Cu data collapse together,
resulting in the approximate source ``size'' as a function of $E_t$. A
weighted average ($\left< Z_0\right>$) of the data
over different exit channels, constructed according to
\begin{equation}
\left< Z_0\right>(E_t) =\sum _Z Zm_Z(E_t)a_Z,
\label{eq:average_size}
\end{equation}
is shown by the symbols in panel c) of Fig.~\ref{fig:Pn}. The weight
$a_Z$ is the standard weight (proportional to the inverse square of
the individual errors).
A similar behavior
is observed in two additional asymmetric reactions $^{129}$Xe+$^{51}$V
and $^{89}$Y (see Fig.~\ref{fig:Pn}d).


The $E_t$ dependence of the source size is tantalizing.
The source size increases quickly to a saturation value.
The fact that $E_t$ is related to impact parameter as well as to the
total excitation energy may explain the observed features.  In the
highly asymmetric reverse kinematic reactions one quickly achieves
sufficient overlap to produce a dominant Xe-like source as one moves
from peripheral to central collisions.


As a special case of the $1/Z$ scaling, the ``saturation'' $m_Z$
values from central collisions of the reverse kinematics reactions
(the top 5\% of the $E_t$ scale, shown by the hatched regions in panel
b) of Fig.~\ref{fig:Pn}), are shown in Fig.~\ref{fig:mZ0_xecu}. The
open symbols represent the scaled quantity $Zm_Z=Z_0$. The solid lines
are weighted averages for the different reactions. The same data in
the form $m_Z/Z_0$ vs. $Z$ are shown by the solid symbols, where the
$1/Z$ dependence is manifested by the good agreement of the data with
the values of $1/Z$ (solid line).
This striking
$1/Z$ dependence of the parameter $m_Z$ is observed for all asymmetric
systems measured.

These overall results for asymmetric reactions suggest the dominance
of a single source, {\em strongly support the hypothesis of uniform
(statistical) exploration of the fragmentation phase space}, and lead
to the interpretation of $Z_0 = Zm_Z$ as the source ``size.'' The
$1/Z$ scaling is general and should be observed in models used to
describe multifragmentation.


In the less asymmetric reactions $^{129}$Xe+$^{197}$Au at 50 and 60
$A$MeV for which at least two sources are plausible, we shall refer
directly to $Z_0=Zm_Z$ as the source size, although we shall see that
now it depends on the fragment $Z$ value as well as on $E_t$. The weak
decrease of the source size with increasing fragment size $Z$, already
observable in $^{129}$Xe+$^{89}$Y (Fig.~\ref{fig:mZ0_xecu}), becomes more
visible in the case of the $^{197}$Au target. At low fragment $Z$
values, the source size $Z_0$ is $\approx70$ and it decreases
monotonically with increasing fragment size $Z$ to to a source size
$Z_0$ of approximately 40-50.  This fragment size dependence seems to
suggest that for the $^{89}$Y target and, most of all, for the
$^{197}$Au target there may be a distribution of sizes, the higher $Z$
fragments being emitted preferentially by the smaller source(s).

The reactions $^{36}$Ar+$^{197}$Au at 35, 50, 80, 110 $A$MeV give a
picture intermediate between the $^{129}$Xe+$^{197}$Au and the
$^{129}$Xe induced reverse kinematics reactions. They also give
information of the source size dependence on bombarding energy. The
source size at low fragment $Z$ increases from $Z_0\approx 30$ to
$Z_0\approx 60$ as the bombarding energy increases, 
$A$MeV, 
while at higher fragment $Z$ the source size increases from
$Z_0\approx20$ to $Z_0\approx40$.

In previous work \cite{Phair95}, it was empirically shown that the
observed charge distributions resulting from nuclear
multifragmentation obey the following invariant form:
\begin {equation}
P_n (Z) \propto \exp\left( {-\frac{B(Z)}{\sqrt{E_t}}-cnZ}\right)
\label {eq:empirical_c}
\end {equation} 
where $n$ is the total intermediate mass fragment 
multiplicity of the event; $E_t$ is the total transverse energy; 
and $B(Z)$ is the ``barrier'' distribution.

From
thermodynamic considerations and percolation simulations, it was shown
that
$c$ in Eq.~(\ref{eq:empirical_c})
vanishes when the gas of IMFs is in equilibrium with a liquid (residue
source, or percolating cluster), and assumes a value $\propto 1/Z_0$
($Z_0$ being the source size) when the source is wholly vaporized
\cite{Moretto97,Phair95,Moretto96}. 

Experimentally, the parameter $c$ undergoes an evolution with
(transverse) energy from approximately zero to a
positive constant \cite{Moretto96}. Thus the source size evolves from
near infinity (an ``infinite'' reservoir of fragments) to the actual
size of the source.  
With the exception of the reactions
$^{129}$Xe+$^{27}$Al,$^{51}$V, Cu, $^{89}$Y, this limit is
attained. Thus, it is possible to plot the value of $Z_0$ determined
from $m_Z$ against that obtained from the $c$ parameter (both for the
top 5\% most central collisions in $E_t$).

\begin{figure}
\centerline{\psfig{file=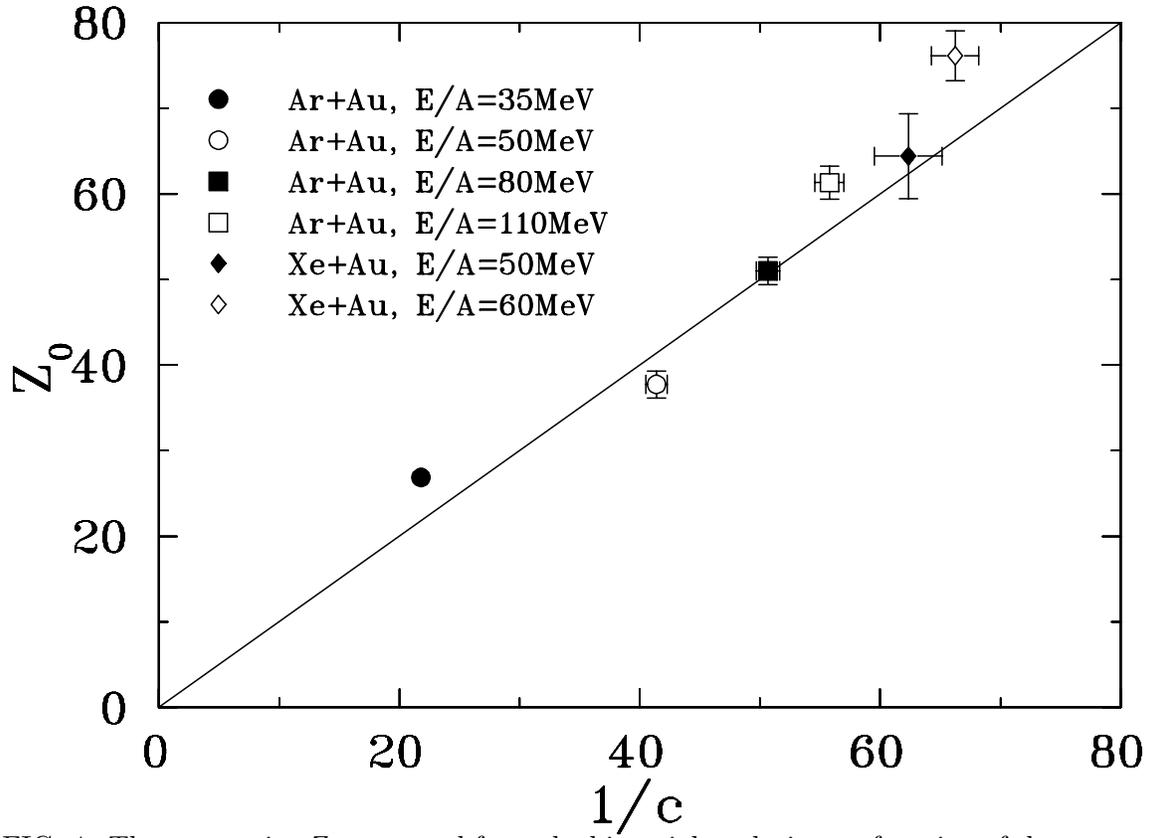,height=15.0cm,angle=90}}
\caption{The source size $Z_0$ extracted from the binomial analysis as
a function of the source size ($1/c$) determined as per
ref.~\protect\cite{Phair95} from central collision of the indicated
reactions.}
\label{fig:m_c}
\end{figure}

Such a plot is shown in
Fig.~\ref{fig:m_c}.  The result is striking. Not only are the two
quantities well correlated, but they also agree quite well 
in absolute value.
This good agreement
gives confidence that we
have gained direct access to the source size. This source (sources) is
specifically the entity that generates the fragments through
``chemical equilibrium''. It does not contain the pre-equilibrium part
which
is often incorporated in other
source reconstruction methods.


In conclusion, we have shown that:
a) the binomial (nearly Poissonian) multiplicity distributions for
individual fragment atomic numbers permit the extraction of the
parameter $m_Z$, the number of throws;
b) $m_Z$ for reactions where a single source is clearly dominant
has the form $m_Z=Z_0/Z$;
c) the $1/Z$ dependence is dramatic proof that the fragmentation
phase space is statistically explored;
d) source size(s) can be extracted and should reflect the
region(s) where chemical (as opposed to physical) equilibrium is
achieved;
e) these source sizes agree with the sizes obtained from the analysis
of multiplicity selected charge distributions in the $E_t$ range where
a single gas phase, or thermodynamic bivariance, prevails.

Acknowledgments

This work was supported by the 
Nuclear Physics Division of
the US Department of Energy, under contract DE-AC03-76SF00098.
One of us (L.B) acknowledges a fellowship from the National Sciences
and Engineering Research Council (NSERC), Canada.

$^{*}$ Present address: Indiana University Cyclotron Facility, 2401
Milo B. Sampson Ln, Bloomington, IN 47408\\


\begin{figure}
\centerline{\psfig{file=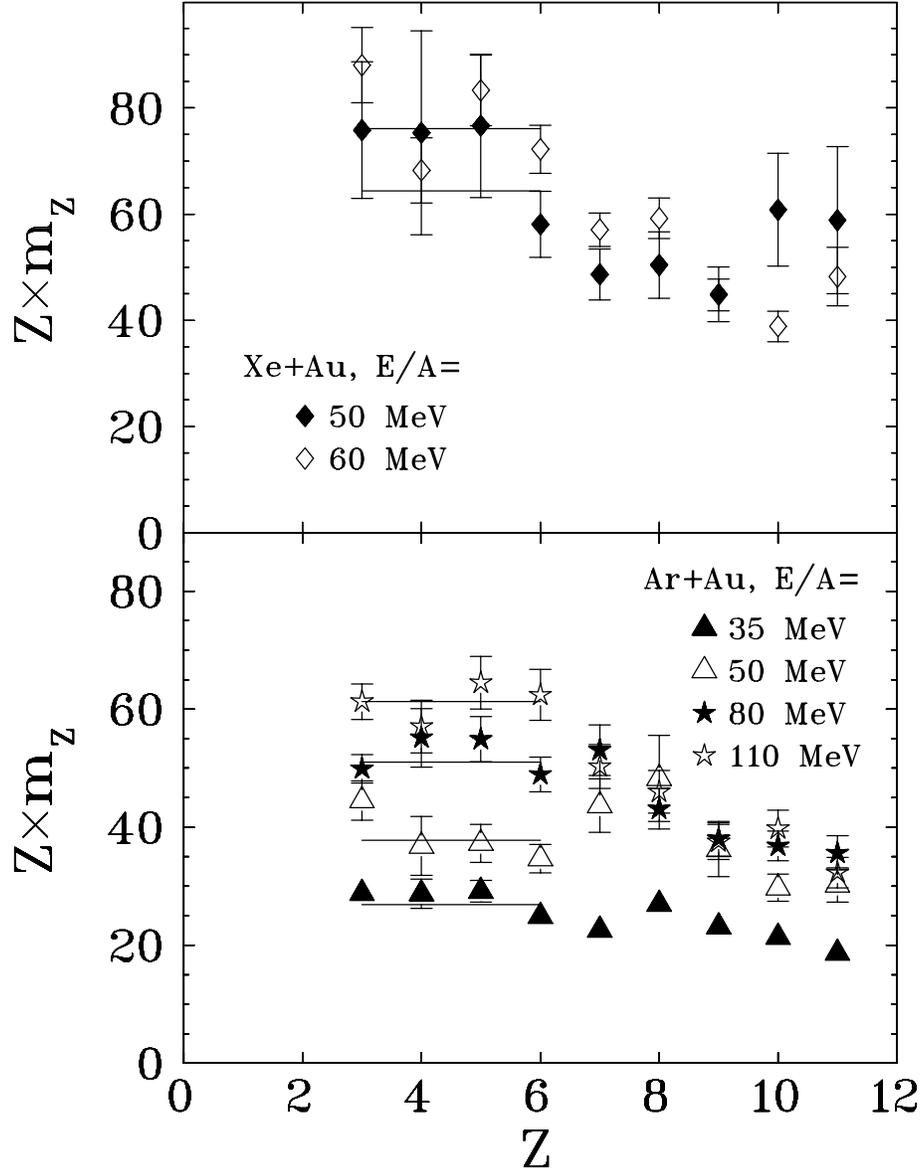,height=15.5cm,angle=0}}
\vspace{0.1cm}
\caption{The scaled values $Z\times m_Z$ are plotted for the indicated
reactions of the 5\% most central collisions. The solid lines
represent weighted averages from $Z=3$ to 6.}
\vspace{.2cm}
\label{fig:mZ0}
\end{figure}

\end{document}